\magnification=1200
\baselineskip=13pt
\overfullrule=0pt
\tolerance=100000
\nopagenumbers

\font\tenbifull=cmmib10 \skewchar\tenbifull='177
\font\tenbimed=cmmib7   \skewchar\tenbimed='177
\font\tenbismall=cmmib5  \skewchar\tenbismall='177
\textfont9=\tenbifull
\scriptfont9=\tenbimed
\scriptscriptfont9=\tenbismall

\mathchardef\alpha="710B
\mathchardef\beta="710C
\mathchardef\gamma="710D
\mathchardef\delta="710E
\mathchardef\epsilon="710F
\mathchardef\zeta="7110
\mathchardef\eta="7111
\mathchardef\theta="7112
\mathchardef\iota="7113
\mathchardef\kappa="7114
\mathchardef\lambda="7115
\mathchardef\mu="7116
\mathchardef\nu="7117
\mathchardef\micron="716F
\mathchardef\xi="7118
\mathchardef\pi="7119
\mathchardef\rho="711A
\mathchardef\sigma="711B
\mathchardef\tau="711C
\mathchardef\upsilon="711D
\mathchardef\phi="711E
\mathchardef\chi="711F
\mathchardef\psi="7120
\mathchardef\omega="7121
\mathchardef\varepsilon="7122
\mathchardef\vartheta="7123
\mathchardef\varphi="7124
\mathchardef\varrho="7125
\mathchardef\varsigma="7126
\mathchardef\varpi="7127

 at 8truept

\

{\hfill \hbox{\vbox{\settabs 1\columns
\+ hep-th/9802070\cr
}}}
\centerline{}
\bigskip
\bigskip
\bigskip
\baselineskip=18pt

\centerline{\bf Non-Local Charges and their Algebra in Topological Field Theory}
\vfill
{\baselineskip=11pt
\centerline{J. C. Brunelli\footnote{*}{e-mail address: brunelli@fsc.ufsc.br}}
\medskip
\medskip
\centerline{Universidade Federal de Santa Catarina}
\centerline{Departamento de F\'\i sica -- CFM}
\centerline{Campus Universit\'ario -- Trindade}
\centerline{C.P. 476, CEP 88040-900}
\centerline{Florian\'opolis, SC -- BRAZIL}
\medskip
\medskip
\centerline{and}
\medskip
\medskip
\centerline{Ashok Das}
\medskip
\medskip
\centerline{Department of Physics and Astronomy}
\centerline{University of Rochester}
\centerline{Rochester, NY 14627 -- USA}
}
\vfill

\centerline{\bf {Abstract}}

\medskip

With the third order Monge-Amp\`ere equation as an example, we show that there exists an infinite number of nonlocal conserved charges associated with the Witten-Dijkgraaf-Verlinde-Verlinde equations. A general prescription for the construction of these charges is given and the charge algebra is calculated bringing out various other interesting features associated with such systems.
\medskip

\vfill
\eject
\headline={\hfill\folio}
\pageno=1

There exists a very interesting connection between 2-d topological field theories (TFT) and integrable equations of hydrodynamic type [1-4]. For example, it is well known that the free energy of the topological sigma model coincides with the Hamiltonian for the isentropic motion of a special one dimensional fluid [4] and that the $A_n$-minimal models can be related to the dispersionless generalized KdV [1] equations and so on. The connection arises primarily from the following observations [4]. In topological field theories [5] (with perturbations), the two point and the three point correlations [6]
$$
\eqalign{
\langle\phi_\alpha\phi_\beta \rangle=&\eta_{\alpha\beta}=\hbox{nondegenerate constant}\cr
\langle\phi_\alpha\phi_\beta\phi_\gamma \rangle=&c_{\alpha\beta\gamma}(t)=
{\partial^3 F(t)\over\partial t^\alpha \partial t^\beta \partial t^\gamma }
\qquad\alpha,\beta,\gamma=1,2,\dots,n
}\eqno(1)
$$
where $t=(t^1,t^2,\dots,t^n)$ (in TFT's these would correspond to coupling constants) and $F(t)$ is the free energy of the theory, define a commutative and associative algebra (with an identity)
$$
e_\alpha e_\beta=c^\gamma_{\alpha\beta}e_\gamma\eqno(2)
$$ 
with $e_\alpha$ defining a basis of the algebra. Conventionally, one defines $e_1=1$ so that the metric
$$
\eta_{\alpha\beta}=\langle\phi_\alpha\phi_\beta \rangle=
{\partial^3 F(t)\over\partial t^1 \partial t^\beta \partial t^\gamma }=
\hbox{constant}\eqno(3)
$$
Furthermore, the two and the three point correlation functions determine all the higher correlation functions. The associativity of the algebra leads to an over-determined set of equations for the free energy
$$
{\partial^3 F(t)\over \partial t^\alpha \partial t^\beta \partial t^\lambda}
\,\eta^{\lambda\mu}
{\partial^3 F(t)\over \partial t^\gamma \partial t^\delta \partial t^\mu}=
{\partial^3 F(t)\over \partial t^\gamma \partial t^\beta \partial t^\lambda}
\,\eta^{\lambda\mu}
{\partial^3 F(t)\over \partial t^\alpha \partial t^\delta \partial t^\mu}
\eqno(4)
$$
These are the Witten-Dijkgraaf-Verlinde-Verlinde (WDVV) equations [5,6] and can be identified with integrable equations of hydrodynamic type [4]. Every solution of the hydrodynamic equation, therefore, leads to a particular solution for the free energy of the TFT.

The simplest, nontrivial equation resulting from Eq. (4) is for $n=3$ (we refer the reader to [4] for details and notations.)
$$
f_{ttt}=f_{xxt}^2-f_{xxx}f_{xtt}\eqno(5)
$$
where $x=t^2$ and $t=t^3$.
This is known as the third order Monge-Amp\`ere equation and has been actively studied in the last few years [7,8]. It is known that this equation has a finite number of conserved charges of hydrodynamic type [7,9], possesses a bi-Hamiltonian structure [8] and can be given a zero curvature description [8]. In this letter, we show that this system possesses an infinite number of nonlocal conserved charges very much like the nonlinear sigma model [10]. However, we also point out differences between the two systems. We calculate the algebra of these charges and bring out other interesting features of this system. We note that even though we choose to work with this simple example of the equations of associativity, the features and the procedures appear to be quite general and would hold for other systems as well.

Equation (5) is known [8] to have a bi-Hamiltonian description in terms of the variables
$$
a=f_{xxx},\quad b=f_{xxt},\quad c=f_{xtt}\eqno(6)
$$
so that
$$
\pmatrix{a\cr
\noalign{\vskip .1truecm}%
b\cr
\noalign{\vskip .1truecm}%
c}_t=
\pmatrix{b\cr 
\noalign{\vskip .1truecm}%
c\cr 
\noalign{\vskip .1truecm}%
b^2-ac}_x=
{\cal D}_1
\pmatrix{
{\delta H_0\over\delta a}\cr
\noalign{\vskip .1truecm}%
{\delta H_0\over\delta b}\cr
\noalign{\vskip .1truecm}%
{\delta H_0\over\delta c}}=
{\cal D}_2
\pmatrix{
{\delta H_2\over\delta a}\cr
\noalign{\vskip .1truecm}%
{\delta H_2\over\delta b}\cr
\noalign{\vskip .1truecm}%
{\delta H_2\over\delta c}}
\eqno(7)
$$
where the Hamiltonian structures and the Hamiltonians have the explicit form $(\partial\equiv {\partial\ \over\partial x})$
$$
\displaylines{
{\cal D}_1=\pmatrix{
-{3\over2}\partial & {1\over 2}\partial a & \partial b\cr
\noalign{\vskip .2truecm}%
{1\over2}a\partial & {1\over 2}(\partial b+b\partial) & \partial c+{1\over2}c\partial\cr
\noalign{\vskip .2truecm}%
b\partial &{1\over2}\partial c+c\partial & 
\matrix{\partial(b^2-ac)\cr+(b^2-ac)\partial}
\cr
},\quad
{\cal D}_2=\pmatrix{
0 & 0 & \partial^3\cr
\noalign{\vskip .2truecm}%
0 & \partial^3 & -\partial^2 a\partial\cr
\noalign{\vskip .2truecm}%
\partial^3 & -\partial a \partial^2 &
\matrix{ \partial^2 b\partial +\partial b \partial^2\cr+\partial a\partial a\partial}
\cr
}\cr\cr
\hfill H_0=\int dx\,c,\quad H_2=-\int dx\left({1\over 2}a(\partial^{-1}b)^2+(\partial^{-1}b)(\partial^{-1}c)\right)\hfill(8)}
$$
It is of interest to note here that the second Hamiltonian is nonlocal. In fact, while supersymmetric integrable systems, in general, possesses an infinite number of nonlocal conserved quantities [11,12], among the bosonic integrable systems it is only the nonlinear sigma model (and the ones to which it can be mapped) that we know which possesses nonlocal charges (again, an infinity of them). It is, therefore worth investigating whether the system of equations (7) possesses more conserved nonlocal charges. As we will show, there does exist an infinity of conserved nonlocal charges for this system.

To begin with, we note that Eq. (7) can be written [8] as a zero curvature condition of the form
$$
\partial_t A-\partial_x B -[A,B]=0\eqno(9)
$$
with
$$
A=\pmatrix{
0 &1 &0\cr
b &a &1\cr
c &b &0\cr},\quad
B=\pmatrix{
0 &0 &1\cr
c &b &0\cr
b^2-ac &c &0\cr}\eqno(10)
$$
In fact, the matrices $A$ and $B$ commute and, therefore one can naively identify
$$
J^{0(0)}=A,\quad J^{1(0)}=-B\eqno(11)
$$
and carry through the standard construction of nonlocal charges as in the case of nonlinear sigma model [13,14]. This naive construction, however, fails to give conserved charges. The cause of the problem is not hard to see. Unlike the nonlinear sigma model, here the matrices $A$ and $B$ do not vanish asymptotically and, therefore, one has to be more careful.
In fact, we only need the spatial part of the current to vanish asymptotically. Therefore, the proper identification of the conserved current at the zeroth order is
$$
J^{\mu(0)}=(A,-B+D)\eqno(12)
$$
with
$$
D=\pmatrix{
0 &0 &1\cr
0 &0 &0\cr
0 &0 &0\cr
}\eqno(13)
$$
so that the zeroth order current can be written consistently as ($\partial^{-1}$ can be identified with the alternating step function)
$$
J^{\mu(0)}=\epsilon^{\mu\nu}\partial_\nu\chi^{(0)},\quad
\chi^{(0)}=(\partial^{-1}J^{0(0)})=(\partial^{-1}A)\eqno(14)
$$
The standard construction of nonlocal charges can now be carried through and the conserved currents at any order can be determined recursively from the zeroth order current (12) as
$$
J^{\mu(n)}=\left(A(\partial^{-1}J^{0(n-1)}),-B(\partial^{-1}J^{0(n-1)})+(\partial^{-1}J^{0(n-1)})D\right)\eqno(15)
$$

Explicitly, with an overall normalization of $1\over2$ for the second and the third densities and neglecting total divergences, we have the first few charge densities (The surface terms, of course, cannot be neglected in the presence of nonlocality. However, in the present case, such terms led to products of lower order charges which are individually conserved and are not fundamentally new charges.)

$$
\displaylines{
{J}^{0(0)}=\pmatrix{
0 & 1 & 0\cr
b & a & 1\cr
c &b & 0\cr
},\quad
{J}^{0(1)}=\pmatrix{
(\partial^{-1}b) &(\partial^{-1}a)  & (\partial^{-1}1)\cr
\noalign{\vskip .1truecm}%
a(\partial^{-1}b)+(\partial^{-1}c) & 0 & -(\partial^{-1}a)\cr
\noalign{\vskip .1truecm}%
0 & -\left(a(\partial^{-1}b)+(\partial^{-1}c)\right)&-(\partial^{-1}b) }\cr\cr\cr
\hfill 
{J}^{0(2)}\!=\!\pmatrix{\!
{1\over 2}\partial^{-1}\left(a(\partial^{-1}b)+(\partial^{-1}c)\right)\! &\!
\matrix{{1\over4}(\partial^{-1}a)^2\cr
-{1\over2}\partial^{-1}(\partial^{-1}b)}\! &\! 
-\partial^{-1}(\partial^{-1}a)\!\cr
\noalign{\vskip .3truecm}%
\!\matrix{{1\over4}b(\partial^{-1}a)^2-{1\over4}(\partial^{-1}b)^2\cr-{1\over2}(\partial^{-1}a)(\partial^{-1}c)}\! &\!
-\partial^{-1}\left(a(\partial^{-1}b)+(\partial^{-1}c)\right)\! & \!
\matrix{{1\over4}(\partial^{-1}a)^2\cr
-{1\over2}\partial^{-1}(\partial^{-1}b)}\!\cr
\noalign{\vskip .3truecm}%
\!\matrix{-{1\over2}a(\partial^{-1}b)^2\cr
-(\partial^{-1}b)(\partial^{-1}c)}\! &\!
\matrix{{1\over4}b(\partial^{-1}a)^2-{1\over4}(\partial^{-1}b)^2\cr
-{1\over2}(\partial^{-1}a)(\partial^{-1}c)}\! & \!
{1\over 2}\partial^{-1}\left(a(\partial^{-1}b)+(\partial^{-1}c)\right)\!\cr
}\hfill\cr
\hfill(16)}
$$
The third order charge density is much more complicated and does not give any new insight. So, we simply note here that the only fundamental new charge that arises in third order has the form
$$
\eqalign{
Q_{21}^{(3)}=\int dx\bigg[{1\over2}&\partial^{-1}\left(c\,\partial^{-1}(\partial^{-1}b)\right)-{1\over2}\partial^{-1}\left((\partial^{-1}b)(\partial^{-1}c)\right)+{1\over4}(\partial^{-1}a)^2(\partial^{-1}c)\cr
-{1\over2}&\partial^{-1}\left(a(\partial^{-1}b)^2\right)-{1\over12}(\partial^{-1}a)^3b-{1\over2}(\partial^{-1}a)\,b\,\partial^{-1}(\partial^{-1}b)\cr
+{1\over2}&b\,(\partial^{-1}b)\,\partial^{-1}(\partial^{-1}a)\bigg]\cr
=-Q_{32}^{(3)}\quad&{}
}\eqno(17)
$$
There are several things to note from Eq. (16). First, the zeroth order charge density leads to the local charges. Second, the lower left corner element of $J^{0(2)}$, namely $\left(J^{0(2)}\right)_{31}$, corresponds precisely to the second Hamiltonian density (8) determined earlier. However, there are now new conserved, nonlocal charges -- in fact, an infinity of them very much like the nonlinear sigma model, but with fundamental differences. First, there does not appear to be any symmetry group associated with these currents unlike in the sigma model. Second, one can check explicitly from the structure in (16) or from the general expression in (15) that it is the lower left $2\times2$ submatrix that gives rise to conserved charges and it appears that there are three fundamentally new charges at every even order and only one at every odd order. Furthermore, even though the other elements do not give rise to conserved charges, in general, the time derivative of the integrals of these elements, except for the $(\ )_{13}$ element, give rise to lower order charges (namely, the second time derivatives of these elements and the third time derivatives of the $(\ )_{13}$ element vanish). This is quite significant for it says that even though they are not conserved, one can form combinations from these elements which are, in fact, conserved. Namely, it is straightforward to check that, at lower orders,
$$
\eqalign{
Q_1^*=&\left(\int b\right)\left(\int\partial^{-1}b\right)-\left(\int c\right)\int\partial^{-1}a\cr
Q_2^*=&Q_{21}^{(1)}\left(\int\partial^{-1} b\right)-\left(\int c\right)
\int\left(-{1\over2}(\partial^{-1}a)^2+\partial^{-1}(\partial^{-1}b)\right)\cr
Q_3^*=&Q_{21}^{(1)}\left(\int\partial^{-1} a\right)-\left(\int b\right)
\int\left(-{1\over2}(\partial^{-1}a)^2+\partial^{-1}(\partial^{-1}b)\right)\cr
Q_4^*=&\left(\int\partial^{-1} b\right)^2-2Q_{31}^{(0)}\int\partial^{-1}(\partial^{-1}a)\cr
}\eqno(18)
$$
and so on are conserved. These charges, indeed, arise in the product of the charge matrices as well as in the algebra of the charges where they are crucial for the closure of the algebra. This is a new feature not found in the nonlinear sigma model. Finally, we note here that the charge matrices are related by the recursion relation in every alternate order as
$$
{\cal D}_1
\pmatrix{
{\delta Q^{(0)}_{ij}\over\delta a}\cr
\noalign{\vskip .1truecm}%
{\delta Q^{(0)}_{ij}\over\delta b}\cr
\noalign{\vskip .1truecm}%
{\delta Q^{(0)}_{ij}\over\delta c}}=
{\cal D}_2
\pmatrix{
{\delta Q^{(2)}_{ij}\over\delta a}\cr
\noalign{\vskip .1truecm}%
{\delta Q^{(2)}_{ij}\over\delta b}\cr
\noalign{\vskip .1truecm}%
{\delta Q^{(2)}_{ij}\over\delta c}},\quad
{\cal D}_1
\pmatrix{
{\delta Q^{(1)}_{ij}\over\delta a}\cr
\noalign{\vskip .1truecm}%
{\delta Q^{(1)}_{ij}\over\delta b}\cr
\noalign{\vskip .1truecm}%
{\delta Q^{(1)}_{ij}\over\delta c}}=
{\cal D}_2
\pmatrix{
{\delta Q^{(3)}_{ij}\over\delta a}\cr
\noalign{\vskip .1truecm}%
{\delta Q^{(3)}_{ij}\over\delta b}\cr
\noalign{\vskip .1truecm}%
{\delta Q^{(3)}_{ij}\over\delta c}}
\eqno(19)
$$
This is a new feature of this model and we conjecture that this holds even at higher orders -- which would suggest that among this infinite set of nonlocal charges, there exists an infinite subset of charges which are in involution. This may explain the origin of integrability in this model considering that it only has a finite number of local charges of hydrodynamic type. It is worth pointing out here that even though the charges are related by the recursion relation in (19), there is no recursion operator since both ${\cal D}_1$ and ${\cal D}_2$ are degenerate (they have zero modes which can be explicitly checked and even follows from the fact that these structures have odd dimensions). Consequently, the construction of higher nonlocal charges through a recursion operator does not work.

We can now ask the interesting question of the algebra of these infinite number of nonlocal charges. It is straightforward to check that the zeroth order charges are all in involution. (We are going to use the first Hamiltonian structure,  ${\cal D}_1$, in evaluating the algebra for simplicity. Similar results will hold with the other structure as well.)
$$
\left\{Q^{(0)}_{ij},Q^{(0)}_{k\ell}\right\}=0\qquad i,j,k,\ell=1,2,3\eqno(20)
$$
This result can be understood simply as follows. Of the three nontrivial charges at this order, $Q^{(0)}_{31}$ and $Q^{(0)}_{21}$ (up to a normalization) correspond respectively to the Hamiltonian and the momentum with respect to ${\cal D}_1$ and, therefore, have vanishing Poisson bracket with all local charges. In fact, by construction, we note that
$$
\left\{Q^{(0)}_{31},Q^{(0)}_{\alpha\beta}\right\}=0
\eqno(21)
$$
when $\alpha,\beta$ are restricted to take values corresponding to the lower left $2\times2$ block.

The calculations get more technical beyond the lowest order and we simply note the results here.
$$
\displaylines{\qquad
\eqalign{
\left\{Q^{(0)}_{21},Q^{(1)}_{21}\right\}=&-{1\over2}Q^{(0)}_{31}\cr
\left\{Q^{(0)}_{22},Q^{(1)}_{21}\right\}=&-{3\over2}Q^{(0)}_{21}\cr
}\hfill(22)}
$$ 
$$ 
\displaylines{\qquad
\eqalign{
\left\{Q^{(0)}_{21},Q^{(2)}_{21}\right\}=&0=
\left\{Q^{(0)}_{21},Q^{(2)}_{31}\right\}=
\left\{Q^{(0)}_{22},Q^{(2)}_{22}\right\}=
\left\{Q^{(0)}_{22},Q^{(2)}_{31}\right\}\cr
\left\{Q^{(0)}_{21},Q^{(2)}_{22}\right\}=&{1\over2}Q^{(1)}_{21}\cr
\left\{Q^{(0)}_{22},Q^{(2)}_{21}\right\}=&-{3\over4}Q^{(1)}_{21}
}\hfill(23)}
$$
$$
\displaylines{\qquad
\eqalign{
\left\{Q^{(0)}_{21},Q^{(3)}_{21}\right\}=&-{1\over2}Q^{(2)}_{31}-{1\over16}Q^{(0)}_{21}Q^{(0)}_{31}Q^{(0)}_{23}-{1\over32}\left(Q^{(0)}_{22}\right)^2Q^{(0)}_{31}\cr
\left\{Q^{(0)}_{22},Q^{(3)}_{21}\right\}=&-{3\over2}Q^{(2)}_{21}-{3\over32}\left(Q^{(0)}_{21}\right)^2Q^{(0)}_{12}\cr
\left\{Q^{(1)}_{21},Q^{(2)}_{21}\right\}=&{1\over2}Q^{(2)}_{31}-{1\over8}Q^{(0)}_{21}Q^{(0)}_{31}Q^{(0)}_{12}+{1\over16}Q^{(0)}_{22}Q^{(0)}_{21}Q^{(0)}_{32}
-{1\over32}\left(Q^{(0)}_{22}\right)^2Q^{(0)}_{31}\cr
\left\{Q^{(1)}_{21},Q^{(2)}_{22}\right\}=&Q^{(2)}_{21}-{1\over16}\left(Q^{(0)}_{22}\right)^2Q^{(0)}_{21}+{1\over16}\left(Q^{(0)}_{21}\right)^2Q^{(0)}_{12}+{1\over8}Q^{(0)}_{22}Q^{(0)}_{23}Q^{(0)}_{31}\cr
\left\{Q^{(1)}_{21},Q^{(2)}_{31}\right\}=&-{1\over16}\left(Q^{(0)}_{21}\right)^3-{1\over8}Q^{(0)}_{21}Q^{(0)}_{22}Q^{(0)}_{31}-{1\over8}\left(Q^{(0)}_{31}\right)^2Q^{(0)}_{12}}\hfill(24)}
$$

Up to this order, the algebra appears to be nonlinear very much like the nonlinear sigma model [15-17]. However, new features arise when we go beyond this order. Before giving higher order results, let us note that the first of the relations in (23) is a manifestation of the recursion relation that we discussed earlier in (19). On the other hand, the first relation in (22) suggests that there is no recursion relation (operator) connecting the charges in the adjacent orders. Let us now record the algebraic relations in the next order.
$$
\displaylines{\qquad
\eqalign{
\left\{Q^{(1)}_{21},Q^{(3)}_{21}\right\}=&-{1\over16}Q^{(1)}_{21}Q^{(0)}_{31}Q^{(0)}_{12}-{1\over16}Q^{(0)}_{21}Q^{*}_{1}\cr
\left\{Q^{(2)}_{21},Q^{(2)}_{22}\right\}=&{1\over2}Q^{(3)}_{21}+{1\over16}Q^{(0)}_{22}Q^{*}_{1}\cr
\left\{Q^{(2)}_{21},Q^{(2)}_{31}\right\}=&-{1\over16}Q^{(0)}_{22}Q^{(0)}_{31}Q^{(1)}_{21}-{1\over32}\left(Q^{(0)}_{21}\right)^2Q^{(1)}_{21}
-{1\over16}Q^{(0)}_{31}Q^{*}_{1}\cr
\left\{Q^{(2)}_{22},Q^{(2)}_{31}\right\}=&{1\over8}Q^{(0)}_{31}Q^{(0)}_{12}Q^{(1)}_{21}-{1\over8}Q^{(0)}_{21}Q^{*}_{1}}\hfill(25)}
$$
As we had noted earlier, the first of the new charges in (18) appears in the charge algebra at this stage and needs to be included for the closure of the algebra. Including these in the set of original charges, we have up to this order the new nontrivial relations in the algebra as
$$
\displaylines{\qquad
\eqalign{
\left\{Q^{(0)}_{21},Q^{*}_{1}\right\}=&-{1\over2}\left(Q^{(0)}_{21}\right)^2+
{1\over2}Q^{(0)}_{22}Q^{(0)}_{31}\cr
\left\{Q^{(0)}_{22},Q^{*}_{1}\right\}=&-{3\over2}Q^{(0)}_{31}Q^{(0)}_{12}\cr
\left\{Q^{(1)}_{21},Q^{*}_{1}\right\}=&-{1\over2}Q^{(0)}_{21}Q^{(1)}_{21}\cr
\left\{Q^{(2)}_{21},Q^{*}_{1}\right\}=&-{1\over2}Q^{(0)}_{21}Q^{(2)}_{21}
+{3\over4}Q^{(0)}_{31}Q^{(2)}_{22}+{1\over32}\left(Q^{(0)}_{21}\right)^3Q^{(0)}_{12}\cr&+{1\over96}\left(Q^{(0)}_{22}\right)^3Q^{(0)}_{31}-{1\over8}Q^{(0)}_{21}Q^{(0)}_{22}Q^{(0)}_{31}Q^{(0)}_{12}\cr
\left\{Q^{(2)}_{22},Q^{*}_{1}\right\}=&-{1\over2}Q^{*}_{2}
+{1\over4}Q^{(0)}_{21}Q^{(0)}_{31}\left(Q^{(0)}_{12}\right)^2
+{1\over16}\left(Q^{(0)}_{22}\right)^2Q^{(0)}_{31}Q^{(0)}_{12}\cr
\left\{Q^{(2)}_{31},Q^{*}_{1}\right\}=&-Q^{(0)}_{21}Q^{(2)}_{31}
+Q^{(0)}_{31}Q^{(2)}_{21}-{1\over16}\left(Q^{(0)}_{21}\right)^2Q^{(0)}_{31}Q^{(0)}_{12}-{1\over16}\left(Q^{(0)}_{22}\right)^2Q^{(0)}_{31}Q^{(0)}_{21}\cr
}\hfill(26)}
$$
$$
\displaylines{\qquad
\eqalign{
\Bigl\{Q^{*}_{1},Q^{*}_{2}\Bigr\}=&-{1\over2}Q^{*}_{3}Q^{(0)}_{31}+
{1\over2}Q^{*}_{4}Q^{(0)}_{31}\cr
&+{1\over16}\left(Q^{(0)}_{22}\right)^2Q^{(0)}_{21}Q^{(0)}_{31}Q^{(0)}_{12}+{1\over8}\left(Q^{(0)}_{21}\right)^2Q^{(0)}_{31}\left(Q^{(0)}_{12}\right)^2\cr
}\hfill(27)}
$$
$$
\displaylines{\qquad
\eqalign{
\left\{Q^{(0)}_{21},Q^{*}_{2}\right\}=&-{1\over2}Q^{(0)}_{21}Q^{(1)}_{21}\cr
\left\{Q^{(0)}_{22},Q^{*}_{2}\right\}=&-{3\over2}Q^{*}_{1}\cr
\left\{Q^{(1)}_{21},Q^{*}_{2}\right\}=&-{1\over2}\left(Q^{(1)}_{21}\right)^2
-{3\over2}Q^{(0)}_{31}Q^{(2)}_{22}+{1\over24}\left(Q^{(0)}_{22}\right)^3Q^{(0)}_{31}\cr&+{3\over8}Q^{(0)}_{22}Q^{(0)}_{21}Q^{(0)}_{31}Q^{(0)}_{12}+{1\over4}\left(Q^{(0)}_{31}\right)^2\left(Q^{(0)}_{12}\right)^2\cr
}\hfill(28)}
$$

Thus, it is clear that new combination charges arise in higher orders. However, the algebra is closed. It is also worth noting here that the algebraic relations in (26)--(28) appear to be of even order unlike the ones in (22)--(25) primarily because of our choice of the new charges as quadratic combinations in (18). Nonlinear algebras are, of course, well know in the study of the nonlinear sigma model [15-17]. As past experience shows [12,17], in such cases, one can redefine the basis of the charges so that the algebra takes the form of a Yangian [18,19]. In the present case, we have tried to redefine the charges to write it as a Yangian. However, we have not been completely successful primarily for two reasons. First, an underlying symmetry structure is lacking and second, the appearance of new combination charges at higher orders makes it even harder. Whether this algebra can, in fact, be written as a Yangian remains an open question 
\bigskip
\leftline{\bf Acknowledgments}
\medskip
 
A. D. would like to thank the members of the Departamento de F\'\i sica at UFSC for hospitality during the period when this work was done. J.C.B. was supported by CNPq, Brazil. A.D. was supported in part by the U.S. 
Department of Energy Grant No. DE-FG-02-91ER40685, NSF-INT-9602559 and a Fulbright grant.  
 \bigskip
\leftline{\bf Appendix}

In this appendix we collect some basic formulae which are useful in calculations with nonlocal terms. By definition
$$
\int\limits_{-\infty}^{+\infty}dx\,(\partial^{-1}A)=\int\hskip-7pt\int\limits_{-\infty}^{+\infty}dx\,dy\,\epsilon(x-y)A(y)\eqno(A1)
$$
where
$$
\epsilon(x-y)=-\epsilon(y-x)= \cases{\phantom{+}{1\over2}&\hbox{ for } $x>y$\cr\noalign{\vskip .3truecm}%
-{1\over2}&\hbox{ for } $x<y$\cr}\eqno(A2)
$$
It follows from this that
$$
\int\limits_{-\infty}^{+\infty}dx\,\partial\left(\prod\limits_{i=1}^{n}(\partial^{-1}A_i)\right)=\cases{0&\hbox{ for $n$ even}\cr
\noalign{\vskip .3truecm}%
{1\over2^{n-1}}\prod\limits_{i=1}^{n}\left(\int A_i\right)&\hbox{ for $n$ odd}\cr}\eqno(A3)
$$
In dealing with nonlocal functions, one should be careful and note that while $\partial\partial^{-1}=1$, it is not true, in general, that $\partial^{-1}\partial=1$. In fact, as can be explicitly checked from the definitions
$$
\int\limits_{-\infty}^{+\infty}dx\,\,\partial^{-1}\partial\prod\limits_{i=1}^{n}(\partial^{-1}A_i)=\cases{\int\prod\limits_{i=1}^{n}(\partial^{-1}A_i)&\hbox{ if $n$ is odd}\cr
\noalign{\vskip .3truecm}%
\int \prod\limits_{i=1}^{n}(\partial^{-1}A_i)
-{1\over2^{n}}\left(\int 1\right)\prod\limits_{i=1}^{n}\left(\int A_i\right)&\hbox{ if $n$ is even}\cr}\eqno(A4)
$$
\medskip
\vfill\eject
\leftline{\bf References}
\bigskip
\item{1.}{I. Krichever, Comm. Math. Phys. {\bf 143}, 627 (1992); Comm. Pure Appl. Math. {\bf 47}, 437 (1994).}
\item{2.}{I. Krichever, in ``New Symmetry Principles in Quantum Field Theory'', ed. J. Fr\"ohlich et al, Plenum Press, New York, 1992.}
\item{3.}{B. Dubrovin and S. Novikov, Russ. Math. Surv. {\bf 44}, 35 (1989).}
\item{4.}{B. Dubrovin,  Nucl. Phys. {\bf B379}, 627 (1993); ``Geometry of 2D Topological Field Theories'', preprint SISSA-89/94FM, SISSA, Trieste (1994), hep-th/9407018.} 
\item{5.}{E. Witten, Nucl. Phys. {\bf B340}, 281 (1990).}
\item{6.}{P. Dijkgraaf, E. Verlinde and H. Verlinde, Nucl. Phys. {\bf 352}, 59 (1991).}
\item{7.}{O. I. Mokhov and E. V. Ferapontov, ``Equations of Associativity in Two-Dimensional Field Theory as Integrable Hamiltonian Nondiagonalizable Systems of Hydrodynamic Type'', preprint, hep-th/9505180.}
\item{8.}{E. V. Ferapontov, C. A. P. Galv\~ao, O. I. Mokhov and Y. Nutku,
Commun. Math. Phys. {\bf 186}, 649 (1997).}
\item{9.}{E. V. Ferapontov, Physica {\bf D63}, 50 (1993).}
\item{10.}{K. Pohlmeyer, Commun. Math. Phys. {\bf 46},107 (1976).}
\item{11.}{P. H. M. Kersten, Phys. Lett. {\bf A134}, 25 (1988); G. H. M. Roelfs and P. H. M. Kersten, J. Math. Phys. {\bf 33}, 2185 (1992); P. Dargis and P. Mathieu, Phys. Lett. {\bf A176}, 67 (1993).}
\item{12.}{J. C. Brunelli and A. Das, Phys. Lett. {\bf B354}, 307 (1995).}
\item{13.}{E. Br\'ezin, C. Itzykson, J. Zinn-Justin and J. B. Zuber,  Phys. Lett. {\bf 82B}, 442 (1979).}
\item{14.}{E. Corrigan and C. K. Zachos, Phys. Lett. {\bf 88B}, 273 (1979); T. L. Curtright and C. K. Zachos, Phys. Rev. {\bf D21}, 411 (1980).}
\item{15.}{H. J. de Vega, H. Eichenherr and J. M. Maillet, Commun. Math. Phys. {\bf 92}, 507 (1984).}
\item{16.}{J. Barcelos-Neto, A. Das and  J. Maharana, Z. Phys. {\bf 30C}, 401 (1986).}
\item{17.}{E. Abdalla, M. C. B. Abdalla, J. C. Brunelli and A. Zadra, Commun. Math. Phys. {\bf 166}, 379 (1994).}
\item{18.}{D. Bernard and A. LeClair, Commun. Math. Phys. {\bf 142}, 99 (1989).}
\item{19.}{N. J. Mackay, Phys. Lett. {\bf B281}, 90 (1992); Phys. Lett. {\bf B308}, 444 (1993).}

\end